\documentclass[aps,prx,reprint,longbibliography]{revtex4-2}
\usepackage{amsmath,amssymb,bm}
\usepackage{graphicx}
\usepackage{physics}
\usepackage{hyperref}
\hypersetup{colorlinks, linkcolor={blue}, citecolor={blue}, urlcolor={blue}}
\usepackage{xcolor}
\newcommand{\bk}{\mathbf{k}}

\newcommand{\bv}{\mathbf{v}}
\newcommand{\jcb}{\mathbf{j}_{c}^{b}}
\newcommand{\gradT}{\boldsymbol{\nabla}T}

\begin{document}

\title{Bichargon Hall Effect}

\author{Ping Tang}
\email{tang.ping.a2@tohoku.ac.jp}
\affiliation{Institute for Materials Research, Tohoku University, Sendai 980-8577, Japan}

\date{\today}
\begin{abstract}
The conventional Hall effect arises from the transverse deflection of
mobile electronic charge carriers in conductors. Here, we predict a bosonic charge Hall effect in charge-ordered insulators mediated by bichargons, collective bosonic
excitations of charge order carrying quantized electric charges $\pm 2e$. In contrast to conventional electronic carriers, a thermally excited
bichargon gas does not support dc electrical conduction in response to an external electric
field despite its
constituents carrying finite electric charges. We show that a temperature gradient can instead drive a diffusive flow of thermal bichargons, while an out-of-plane magnetic
field deflects the oppositely charged bichargons in opposite transverse
directions via the Lorentz force, leading to a finite Hall charge current and
opposite charge accumulations at the sample edges. Our results broaden the paradigm of charge Hall transport beyond conventional
electronic conductors, establishing that a charge Hall response can emerge in
an electrically \emph{insulating} system through charged bosonic quasiparticles.
\end{abstract}

\maketitle

\textit{Introduction.---}The Hall effect is one of the most fundamental transport phenomena in
condensed matter. The Hall response originally refers to the transverse
deflection of electronic charge carriers  in a conductor by the
Lorentz force under a magnetic field, giving rise to a transverse charge
current and Hall voltage~\cite{Hall1879,AshcroftMermin}. Hall-type phenomena have since been extended beyond electronic charge transport in conductors to bosonic
quasiparticles that transport quantities other than electric charge. Since bosonic
quasiparticles do not require an electrically conducting host, their Hall
responses can arise in insulators, where quasiparticle transport can instead
be driven by a temperature gradient. For example, phonons contribute to a thermal
Hall effect in dielectrics, whereby a longitudinal temperature gradient
generates a transverse heat current~\cite{Strohm2005,inyushkin2007phonon}. Magnons, the quanta of spin-wave excitations,
carry spin angular momentum and can exhibit a magnonic spin Hall effect in
magnetic insulators~\cite{Cheng2016,Zyuzin2016}, analogous to the electronic spin Hall effect in
conductors~\cite{RevModPhys.87.1213}. Analogously, ferrons~\cite{Tang2022,Bauer2022,PhysRevApplied.20.050501,Tang2024,
choe2026observation,zhang2026electric,shen2025observation,itoh2026observation}, quasiparticle excitations of ferroelectric
order that carry an intrinsic electric dipole moment, have recently been proposed to
generate a transverse electric-dipole current in response to a temperature
gradient in ferroelectric insulators~\cite{lopez2026ferron}. Unlike the ordinary Hall effect
of electronic carriers, however, these bosonic Hall phenomena involve
charge-neutral quasiparticles and generally arise from extrinsic asymmetric scattering~\cite{Sheng2006,Kagan2008,PhysRevLett.113.265901} or geometric effects associated with Berry curvature~\cite{PhysRevLett.105.225901,PhysRevB.86.104305,Cheng2016,Zyuzin2016}, rather than directly from the electromagnetic Lorentz force. While charge transport is conventionally associated with (semi)conductors, it remains a fundamental open question whether a bosonic counterpart of the charge Hall effect can emerge in an electrically \emph{insulating} system. 

\begin{figure}
    \centering
    \includegraphics[width=7.2 cm]{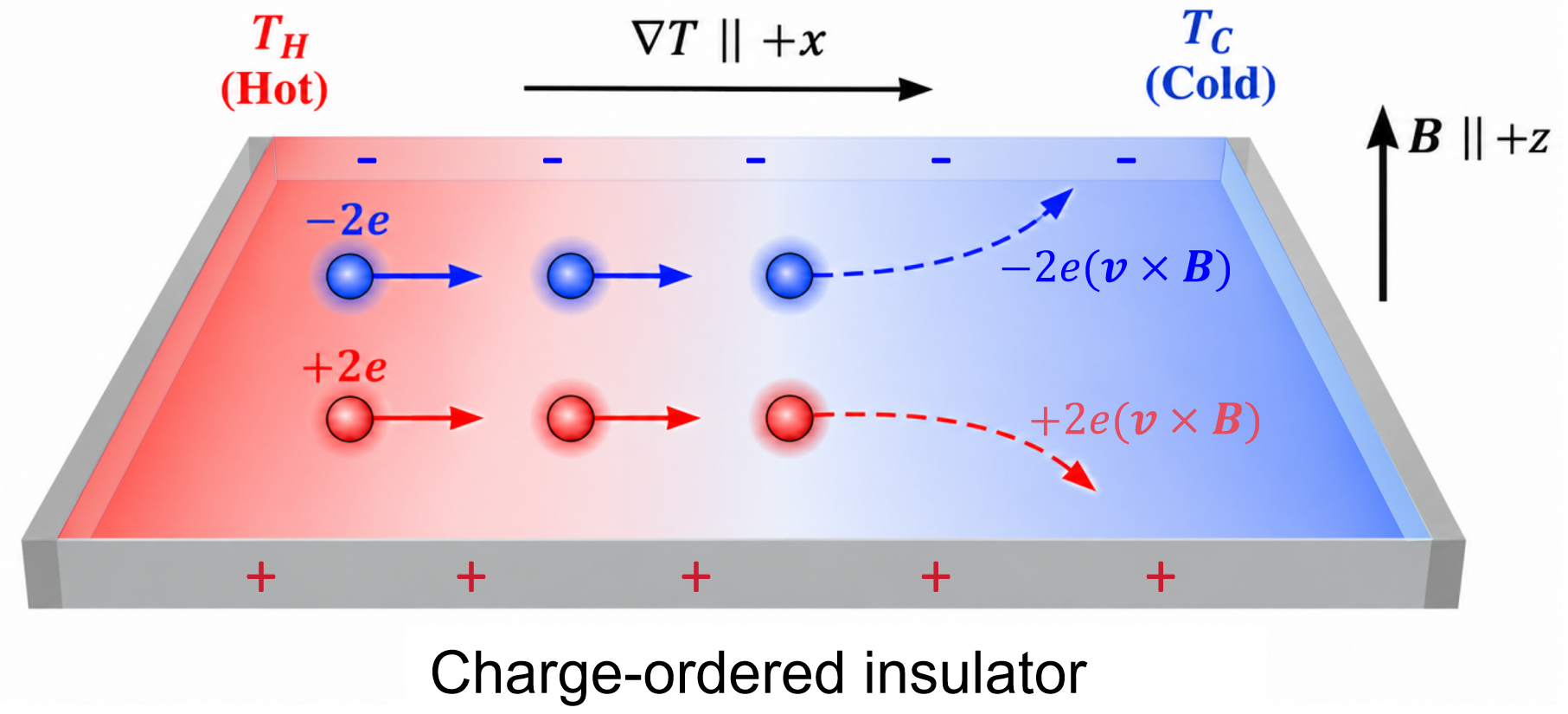}
    \caption{Schematic of the bichargon Hall effect in a charge-ordered
insulator.~Two branches of bichargons carrying opposite electric
charges $\pm 2e$ diffuse in the same direction without generating a net
longitudinal charge current under an applied temperature gradient, while an
out-of-plane magnetic field deflects them in opposite transverse directions
via the Lorentz force, leading to a finite Hall charge current and thereby
opposite charge accumulations at the two lateral edges.}
    \label{fig-1}
\end{figure}

Charge order, the spatial organization of electronic charge, is ubiquitous in strongly correlated materials, including transition-metal oxides~\cite{Attfield2006,Senn2012} and organic molecular conductors~\cite{Takahashi2006,Seo2006}. A minimal microscopic description
of charge order is provided by the extended Hubbard model~\cite{PhysRevLett.53.2327,PhysRevB.39.9397,PhysRevLett.106.236805,PhysRevLett.110.166401,PhysRevLett.111.036601,PhysRevB.95.115149,PhysRevB.99.245146}, which incorporates
intersite Coulomb repulsion in addition to the on-site interaction. In the strong charge-ordered regime, the low-energy electronic degrees of freedom can be mapped onto an effective pseudospin model~\cite{tang2026pseudospin}, establishing a close correspondence between charge- and spin-ordered systems. While magnons in spin-ordered systems carry spin angular momentum,
the collective excitations of staggered charge order can form two branches of
bosonic quasiparticles termed \emph{bichargons}~\cite{tang2026collective}, which carry
opposite quantized electric charges $\pm 2e$. Remarkably, despite carrying electric charge, a thermal bichargon gas does not support dc electrical conduction in response to a static electric field~\cite{tang2026collective}, in contrast to conventional electronic charge carriers, thereby preserving the electrically insulating character of the charge-ordered state. Charge-ordered
insulators thus provide a natural platform for realizing a bosonic charge Hall
effect. 

In this Letter, we predict a bichargon Hall effect in a charge-ordered
insulator mediated by two oppositely charged bichargon branches.~Using the semiclassical
Boltzmann equation, we show that a longitudinal temperature gradient can drive
the diffusion of both bichargon branches in the same direction, while an
out-of-plane magnetic field deflects the oppositely charged bichargons toward
opposite transverse directions. Although the longitudinal thermoelectric charge currents carried by the two branches
cancel, their transverse Hall charge currents add, leading to an
antisymmetric bichargon charge accumulation at the lateral edges of a finite
sample, as illustrated in Fig.~\ref{fig-1}. The charge-compensated bichargon gas therefore exhibits a finite
transverse charge Hall response despite the absence of longitudinal
thermoelectric charge transport. The bichargon Hall effect thus extends
charge Hall transport from conventional electronic carriers to charged
bosonic collective excitations and provides a direct transport signature of
bichargons in charge-ordered systems.


\begin{figure}[t]
    \centering
    \includegraphics[width=8.2 cm]{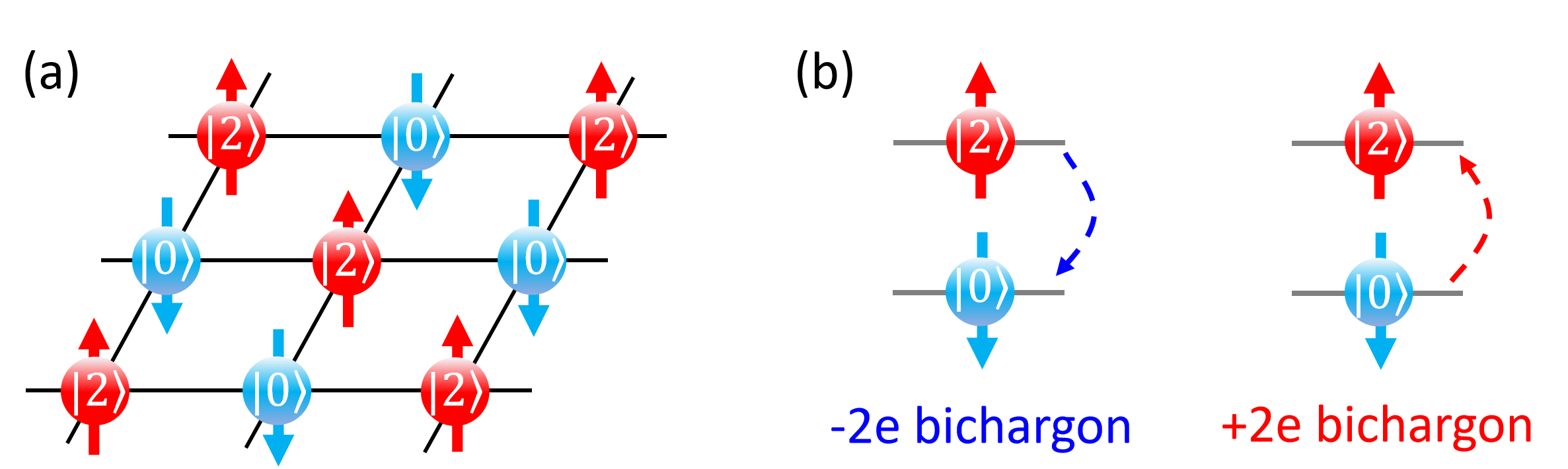}
    \caption{(a) Checkerboard charge order in the half-filled extended Hubbard model on a
two-dimensional square lattice, where the alternating empty $|0\rangle$ and
doubly occupied $|2\rangle$ sites can be mapped onto staggered pseudospin-$1/2$
order. (b) Two branches of oppositely charged ($\pm 2e$) bichargons
associated with pseudospin-flip excitations, analogous to the two magnon
branches in a bipartite antiferromagnet carrying opposite spin angular
momenta $\pm\hbar$. The two branches of bichargon excitations are energetically degenerate and equally populated in equilibrium, thereby leaving the total electric charge
of the system unchanged.}
    \label{fig-2}
\end{figure}

\textit{Model.---}We consider a minimal extended Hubbard model at half filling on a
two-dimensional square lattice,
\begin{align}
H=&-t\sum_{\langle ij\rangle,\sigma}\left(c_{i\sigma}^{\dagger}c_{j\sigma}+\mathrm{H.c.}\right)
-\mu\sum_i n_i \nonumber\\&+U\sum_i n_{i\uparrow}n_{i\downarrow}+V\sum_{\langle ij\rangle}n_i n_j,
\label{eq:Hubbard}
\end{align}
where $c_{i\sigma}^{(\dagger)}$ annihilates (creates) an electron with spin
$\sigma$ at site $i$, $n_{i\sigma}=c_{i\sigma}^{\dagger}c_{i\sigma}$, and
$n_i=n_{i\uparrow}+n_{i\downarrow}$. Here, $t$ denotes the nearest-neighbor
hopping amplitude, $\mu$ the chemical potential, and $U>0$ and $V>0$ the on-site
and nearest-neighbor Coulomb repulsions, respectively. At half filling, $U$ favors a singly occupied Mott state with
antiferromagnetic correlations, whereas $V$ favors a
checkerboard charge-ordered state with alternating empty and doubly occupied
sites. In the atomic limit ($t=0$), the two states have energies
$E_{\mathrm{Mott}}=ZV/2$ and $E_{\mathrm{CO}}=U/2$ per site, respectively,
so that the charge-ordered state is energetically favored for $V>U/Z$~\cite{PhysRevB.3.2662,wolff1983saddle,PhysRevB.48.7140}, where $Z=4$ is the coordination number of the square
lattice. Here we focus on the charge-ordered regime at the particle-hole-symmetric half-filled point, $\mu=U/2+ZV$, with $U,V\gg t$ and $V$ well above the critical value
$V_{\mathrm{CO}}$, where
the ground state exhibits robust insulating checkerboard charge
order~\cite{PhysRevB.95.115149,PhysRevB.97.115117}, as illustrated in Fig.~\ref{fig-2}. In this regime, the empty and doubly occupied states,
$|0\rangle_i$ and $|2\rangle_i$, respectively, span the low-energy Hilbert
space at each site and define a local pseudospin-$1/2$ degree of freedom:
\begin{equation}
\tau_i^z=\frac{n_i-1}{2},
\qquad
\tau_i^+=c_{i\uparrow}^{\dagger}c_{i\downarrow}^{\dagger},
\qquad
\tau_i^-=c_{i\downarrow}c_{i\uparrow},
\label{pse}
\end{equation}
where the longitudinal component $\tau_i^z$
distinguishes the doubly occupied and empty states, with eigenvalues $+1/2$
and $-1/2$, respectively, whereas the ladder operators $\tau_i^\pm$ describe
coherent electron-pair transitions between the two local charge states, $|0\rangle_i\leftrightarrow|2\rangle_i$. Projecting
Eq.~(\ref{eq:Hubbard}) onto this low-energy subspace yields the effective
pseudospin Hamiltonian~\cite{tang2026pseudospin}
\begin{equation}
H_{\rm eff}
=
\sum_{\langle ij\rangle}
\left[
J_z\tau_i^z\tau_j^z
-
\frac{J_\perp}{2}
\left(
\tau_i^+\tau_j^-+\tau_i^-\tau_j^+
\right)
\right]
-h_z\sum_i\tau_i^z ,
\label{eq:Hps}
\end{equation}
where $J_{z}\simeq 4V+\mathcal{O}(t^2/\Delta)$ and $J_{\perp}\simeq4t^2/\Delta$ represent the Ising-type and transverse pseudospin
exchange interactions between neighboring sites, respectively, with $\Delta\simeq (2Z-1)V-U$ denoting the energy cost of creating virtual intermediate singly occupied states on bond $\langle ij\rangle$ by electron hopping. The uniform longitudinal pseudospin field
$h_z=2\mu-U-2ZV$ vanishes at the particle-hole-symmetric half-filled point $\mu=U/2+ZV$. For $J_z>J_\perp>0$, Eq.~(\ref{eq:Hps}) favors a ground
state with staggered pseudospin order, corresponding to checkerboard charge
order in which doubly occupied ($\tau_i^z=+1/2$) and empty
($\tau_i^z=-1/2$) sites alternate between the two sublattices. Assuming the equilibrium staggered pseudospin configuration
$\langle\tau_A^z\rangle=\tau_0$ and
$\langle\tau_B^z\rangle=-\tau_0$, where $A$ and $B$ label the two
sublattices, we introduce the linearized Holstein--Primakoff transformation
$\tau_{iA}^{z}=\tau_0-a_i^\dagger a_i$,
$\tau_{iA}^{+}\simeq\sqrt{2\tau_0}\,a_i$,
$\tau_{jB}^{z}=-\tau_0+b_j^\dagger b_j$, and
$\tau_{jB}^{+}\simeq\sqrt{2\tau_0}\,b_j^\dagger$,
where $a_i^{(\dagger)}$ and $b_j^{(\dagger)}$ are bosonic operators
describing small pseudospin fluctuations on the $A$ and $B$ sublattices,
respectively. After performing Fourier and Bogoliubov transformations,
Eq.~(\ref{eq:Hps}) reduces to the quadratic Hamiltonian for the two
degenerate bichargon branches,
\begin{equation}
H_{\rm eff}
=
\sum_{\bk}\varepsilon_{\bk}\left(
\alpha_{\bk}^{\dagger}\alpha_{\bk}
+\beta_{\bk}^{\dagger}\beta_{\bk}
\right)+\text{const.},
\end{equation}
where $\varepsilon_{\bk}
=
Z\tau_{0}\sqrt{J_z^2-J_{\perp}^2\gamma_{\bk}^2}$ is the dispersion, with
$\gamma_{\bk}=(\cos k_xa+\cos k_ya)/2$ the lattice structure factor
and $a$ the lattice constant. In the long-wavelength limit, $\varepsilon_{\mathbf{k}}=\sqrt{\Delta_{b}^2+\hbar^2c_{b}^2k^2}$, with $\Delta_{b}=Z\tau_0\sqrt{J_z^2-J_\perp^2}$ the bichargon excitation gap and $c_{b}=2\sqrt{2}\tau_{0}J_{\perp}a/\hbar$ the characteristic velocity. In close analogy with the two magnon branches of a uniaxial
antiferromagnet carrying opposite spin angular momenta
$\pm\hbar$~\cite{rezende2020fundamentals,rezende2019introduction}, the two
bichargon branches carry opposite quantized electric charges,
$q_\alpha=+2e$ and $q_\beta=-2e$. This follows directly from the resultant variation in the total
electric charge relative to the half-filling,
\begin{equation}
Q
\equiv
-e\sum_i(n_i-1)
=
2e\sum_{\bk}
\left(
\alpha_{\bk}^{\dagger}\alpha_{\bk}
-
\beta_{\bk}^{\dagger}\beta_{\bk}
\right),
\end{equation}
where $ e>0$ is the elementary charge. Since the two oppositely charged branches are degenerate and equally
populated in thermal equilibrium, their contributions to $Q$ exactly cancel,
such that the system remains charge neutral despite the
presence of a thermal bichargon gas. For the bichargon Hall effect studied
below, we focus on a charge-ordered insulator, where the two bichargon branches are appreciably populated, while
conventional electronic carriers remain thermally suppressed by a relatively
large electronic band gap.

\textit{Bichargon Hall response.---}We consider a two-dimensional thermal bichargon gas subject to an in-plane
temperature gradient $\boldsymbol{\nabla}T$ and electric field $\mathbf{E}$,
as well as an out-of-plane magnetic field $\mathbf{B}=B\hat{\bm z}$. The semiclassical Boltzmann equation
for bichargons within the relaxation-time approximation reads
\begin{equation}
\mathbf{v}\cdot\nabla_{\mathbf{r}} f_{\sigma}
+\frac{q_{\sigma}}{\hbar}(\mathbf{E}+\mathbf{v}\times\mathbf{B}) \cdot\boldsymbol{\nabla}_{\mathbf{k}}f_{\sigma}
=-\frac{f_{\sigma}-f_{0}}{\tau},
\label{eq:boltzmann}
\end{equation}
where $f_{\sigma}(\mathbf{r},\mathbf{k})$ is the distribution function of
bichargons in branch $\sigma$ with electric charge $q_{\sigma}$, and
$\mathbf{v}=\partial\varepsilon_{\mathbf{k}}/\partial(\hbar\mathbf{k})$
is the group velocity, which is identical for the two branches. The first term on the left-hand side of Eq.~(\ref{eq:boltzmann}) describes
spatial diffusion due to an inhomogeneous distribution, whereas the
second represents momentum-space drift under the electric and Lorentz forces, with $\hbar\dot{\mathbf{k}}
=q_{\sigma}(\mathbf{E}+\mathbf{v}\times\mathbf{B})$. The right-hand side accounts for relaxation with a constant relaxation
time $\tau$ toward the local-equilibrium Bose--Einstein distribution
\begin{equation}
    f_{0}(\bar{\varepsilon}_{\mathbf{k}})=\left\{\exp\left(\frac{\bar{\varepsilon}_{\mathbf{k}}(\mathbf{r})}{k_{B}T(\mathbf{r})}\right)-1\right\}^{-1},
\end{equation}
where $\bar{\varepsilon}_{\mathbf{k}}(\mathbf{r})
=\varepsilon_{\mathbf{k}}
-q_{\sigma}\mathbf{r}\cdot\mathbf{E}$
is the local bichargon energy in the electric field. In contrast to electrons, bichargons are
nonconserved quasiparticles with a \textit{vanishing} equilibrium chemical potential. An applied electric field therefore
not only directly drives bichargon drift through the electric force
$q_\sigma\mathbf E$, but also induces a spatially inhomogeneous
local-equilibrium distribution through the position-dependent electrostatic
energy. 

For an isotropic band, $\varepsilon_{\bk}=\varepsilon_{k}$,
solving Eq.~(\ref{eq:boltzmann}) to linear order in the temperature gradient
yields
\begin{align}
f_{\sigma}
= &f_0(\bar{\varepsilon}_{\mathbf{k}})
-
\frac{\tau}
{1+\left(\omega_{c}\tau\right)^2}
\frac{\partial f_{0}(\bar{\varepsilon}_{\mathbf{k}})}{\partial T}
\nonumber\\
&\times \bv\cdot\left[
\gradT-\frac{q_{\sigma}}{\vert q_{\sigma}\vert}
\omega_{c}\tau(\hat{\bm z}\times\gradT)
\right]
\label{eq:deltaf}
\end{align}
where $\omega_c(k)=2eB/m_b(k)$
is the bichargon cyclotron frequency, with
$m_b(k)=\hbar^2k(\partial\varepsilon_k/\partial k)^{-1}$
the cyclotron mass of the isotropic bichargon band. The electric-field-driven drift is exactly compensated by diffusion
arising from the field-induced spatial redistribution of bichargons. Consequently, the electric field enters only through the local-equilibrium
distribution $f_0(\bar{\varepsilon}_{\mathbf{k}})$ and generates no
nonequilibrium transport correction, i.e., the thermal bichargon gas exhibits
no dc electrical conduction despite carrying electric charge, in sharp
contrast to conventional electronic charge carriers. A bichargon charge current can nevertheless be driven by a temperature
gradient,
\begin{equation}
\jcb\equiv
\sum_{\sigma}
\int\frac{d^2\bk}{(2\pi)^2}
q_\sigma\bv f_{\sigma}=\mathcal{L}_{H}^{b}\,\hat{\bm z}\times\gradT
\label{jc}
\end{equation}
where $\mathcal{L}_{H}^{b}$ represent
the transverse (Hall) thermoelectric coefficient
of the bichargons. Since the two bichargon branches are degenerate and equally populated, their
longitudinal thermoelectric currents cancel because of their opposite charges. In contrast, the Lorentz force
deflects the oppositely charged bichargons in opposite transverse directions,
causing their transverse charge currents to add, as illustrated in
Fig.~\ref{fig-1}. The resulting Hall thermoelectric
coefficient is
\begin{align}
\mathcal{L}_{H}^{b}
=& 
\sum_\sigma
\int\frac{d^2\bk}{(2\pi)^2}
\frac{q_\sigma^2
\tau^2 v^2 B
}{
2m_{b}[1+(\omega_{c}\tau)^2
]}
\frac{\partial f_{0}}{\partial T}\label{LH}.
\end{align}
For the relativistic bichargon dispersion
$\varepsilon_k\simeq
\sqrt{\Delta_b^2+\hbar^2c_b^2k^2}$, Eq.~(\ref{LH}) reduces,
in the weak-field limit $\omega_c\tau\ll1$ and for
$k_BT\gtrsim\Delta_b$, to
\begin{equation}
\mathcal{L}_{H}^{b}\simeq
\frac{2e^2\tau^2 c_{b}^2 B}{\pi\hbar^2}
\left(\ln\frac{k_{B}T}{\Delta_{b}}+\frac{1}{2}\right),
\label{eq:weakfield}
\end{equation}
which is linear in $B$ and logarithmically enhanced as the bichargon gap
decreases. This transverse thermoelectric response bears a close analogy
to the ambipolar Nernst response of electrons and holes in two-carrier
electronic systems~\cite{Price1955,Delves1965,Bel2003}, with the
crucial distinction that here the charge carriers are bosonic collective
excitations of an electrically insulating charge-ordered state. Thus, a finite Hall thermoelectric response together with vanishing
longitudinal charge transport provides a characteristic transport signature
of bichargons in charge-ordered insulators.

\textit{Edge accumulation.---}For a temperature gradient applied along $x$ to a strip of
charge-ordered insulator of width $w$ along $y$, the resulting Hall
current leads to a bichargon charge accumulation $\rho_{c}^{b}(y)$ near
the sample edges. Since bichargons are bosonic excitations of the charge-ordered state,
their quasiparticle number is not conserved and can relax through coupling
to other degrees of freedom. Analogous to magnon spin diffusion in magnetic insulators
~\cite{PhysRevLett.109.096603,PhysRevB.94.014412,PhysRevB.93.054412},
the bichargon charge accumulation obeys the diffusion-relaxation equation
$\partial_y^2\rho_c^b=\rho_c^b/\lambda_b^2$, where $\lambda_b$ is the phenomenological bichargon diffusion length. The limit
$\lambda_b\rightarrow\infty$ corresponds to the absence of bichargon-charge
relaxation; a microscopic calculation of $\lambda_b$ requires accounting for the
relaxation channels of bichargons, including their coupling to electronic
degrees of freedom within the system and its surroundings. The resulting inhomogeneous bichargon
accumulation, on the other hand, generates a backflow diffusion current,
modifying the transverse charge current to
\begin{equation}
\mathbf{j}_{c}^{b}\cdot\hat{\mathbf{y}}
=\mathcal{L}_{H}^{b}\partial_x T
-\mathcal{D}_b\partial_y\rho_c^b,
\label{jy}
\end{equation}
where $\mathcal{D}_{b}=\tau\langle v^2\rangle/2$ is the bichargon
diffusion coefficient, with $\langle v^2\rangle$ the mean-square
velocity. Imposing the zero-current boundary condition at $y=\pm w/2$ yields
\begin{equation}
\rho_c^{b}(y)
=
\frac{\mathcal{L}_{H}^{b}\lambda_b}{\mathcal{D}_b}
\frac{
\sinh(y/\lambda_b)
}{
\cosh(w/2\lambda_b)
}
\partial_x T,
\label{eq:accum}
\end{equation}
which exhibits an antisymmetric distribution across the strip over the
characteristic length scale $\lambda_b$. The resulting transverse charge
separation generates an electrostatic potential difference between the two
edges. The polarity of this potential difference reverses upon reversing
either the magnetic field or the temperature gradient, providing an electrical signature of
the bichargon Hall effect. 

We finally estimate the order of magnitude of the bichargon Hall response
using representative model parameters $J_z=10~\mathrm{meV}$,
$J_\perp=6~\mathrm{meV}$, and $a=5\,$\AA. These parameters yield a bichargon excitation gap $\Delta_b\simeq16~\mathrm{meV}$ and a characteristic velocity
$c_b\simeq6.4\times10^3~\mathrm{m/s}$, both lying within the
characteristic scales of collective charge excitations observed in charge-ordered systems~\cite{PhysRevB.66.024401,Ren2004}.
For a relaxation time $\tau=1\,\mathrm{ps}$, we obtain
$\mathcal{L}_H^b/B\simeq8.9\times10^{-10}\,$A/(KT) at room temperature, which is more than one
order of magnitude larger than the corresponding electronic response in
conventional nonmagnetic metals~\cite{notemetal}. This enhancement can be attributed in part to the absence of Sondheimer
cancellation for the bichargon Hall response, which strongly suppresses the Nernst response of conventional electronic carriers~\cite{Sondheimer1948}. The large magnitude of the bichargon
Hall response should facilitate its experimental detection.

\textit{Discussion and conclusion.---}The bichargon Hall effect differs from the Hall transport of charge-neutral
bosonic quasiparticles such as magnons and phonons, which typically relies on
Berry curvature or asymmetric scattering. Bichargons, by contrast, couple
directly to a magnetic field through the ordinary Lorentz force and thereby
generate a transverse charge response, as do conventional electronic charge
carriers. Remarkably, the bichargon Hall effect does not require electrical
conduction and can occur in electrically \emph{insulating} charge-ordered
systems without carrier doping. In charge-ordered insulators, however, thermally excited intrinsic electronic
carriers (electrons and holes) may also contribute to a transverse charge
response in the presence of an external magnetic field. A charge-ordered system with a pronounced separation of energy scales
between the electronic band gap $E_g$ and the bichargon excitation gap
$\Delta_b$ therefore provides an experimentally favorable platform for
distinguishing the bichargon Hall effect within the temperature window
$\Delta_b \lesssim k_B T \ll E_g$, where bichargons are appreciably populated
while electron--hole excitations remain exponentially suppressed. A transverse thermoelectric charge response in the absence of appreciable conventional electrical conductivity would therefore signal a dominant bichargon Hall contribution over thermally activated electronic transport. Promising candidate materials include quasi-two-dimensional organic
charge-transfer salts, such as
$\beta$-(meso-DMBEDT-TTF)$_2$PF$_6$~\cite{Kimura2006},
and half-doped layered transition-metal oxides, such as
La$_{1.5}$Sr$_{0.5}$NiO$_4$~\cite{Kajimoto2003}, both of which
exhibit insulating checkerboard-type charge-ordered states. Although their
microscopic lattices and fillings differ from those of the minimal half-filled
square-lattice model considered here, they provide promising platforms
combining strong electronic charge disproportionation, insulating behavior,
and low dimensionality.

In conclusion, we predict a sizable charge Hall response in charge-ordered insulators mediated by bichargons, the bosonic collective quasiparticle excitations
of charge order. Under a temperature gradient, the two oppositely charged bichargon
branches diffuse in the same longitudinal direction but are deflected toward
opposite transverse directions by the Lorentz force in an applied magnetic
field. Their longitudinal charge currents cancel, whereas their transverse
charge currents add, producing a finite Hall charge current and antisymmetric
edge charge accumulation. The bichargon Hall effect provides a bosonic counterpart
of the conventional Hall effect of electronic charge carriers in conductors,
yet remarkably persists in the absence of electrical conduction. Our findings thus extend the paradigm of charge Hall transport beyond
conventional electronic conductors to electrically insulating systems, where
the Hall response is mediated by charged bosonic quasiparticles. Beyond the
fundamental significance, the bichargon Hall effect opens possibilities for
exploiting transverse thermoelectric functionality in electrically insulating systems.


\textit{Acknowledges.}---This work was supported by JSPS KAKENHI Grant-in-Aid for Scientific Research (B) (Grant No. 26K00625).

\bibliography{reference}
\end{document}